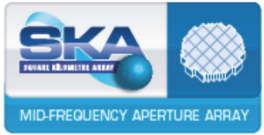

# MANTIS:
# The Mid-Frequency Aperture Array Transient and Intensity-Mapping System

W.A. van Cappellen[1] on behalf of the AAMID Consortium

Contributions and support from M. Santos[2,3,4], J.P. Macquart[5], F. Abdalla[6,7], E. Petroff[1], A. Siemion[1,8,10], R. Taylor[2,10], O. Smirnov[3,7], D. Davidson[11], J. Broderick[1], J. van Leeuwen[1,12], P. Woudt[10], M.A. Garrett[13,14,1], A.J. Faulkner[15], G. Kruithof[1], S.A. Torchinsky[16], I.M. van Bemmel[17] and J. Hessels[1,12].

*Affiliations can be found after the references*

## Summary

The objective of this paper is to present the main characteristics of a wide-field MFAA precursor that we envisage to be built at the SKA site in South Africa. Known as MANTIS (the Mid-Frequency Aperture Array Transient and Intensity-Mapping System), this ambitious instrument will represent the next logical step towards the MFAA based SKA telescope. The goal is to use innovative aperture array technology at cm wavelengths, in order to demonstrate the feasibility of deploying huge collecting areas at modest construction and operational cost. Such a transformative step is required in order to continue the exponential progress in radio telescope performance, and to make the ambitious scale of the SKA Phase 2 a realistic near-time proposition.

This paper summarizes the ideas that were discussed at the 2016 MFAA/MIDPREP workshop in Cape Town. Consultations with the science community at workshops in Stellenbosch (2014) and Cape Town (2016) indicated a strong interest in a wide-field science demonstrator instrument with a collecting area of about 1500 – 2500 $m^2$, operating in the 450 – 1450 MHz frequency range. Such an instrument could perform an outstanding science programme in its own right but would also support and provide leverage to various MeerKAT and SKA1-Mid science opportunities. An instrument of this size matches very well with the necessity to realize a sizable system to demonstrate competitiveness, feasibility and technology readiness of several key MFAA technologies and concepts, as required for the MFAA PDR. MANTIS will also provide a new reference for the costing of an MFAA based SKA, both the deployment costs (hardware and installation) and the operational costs (in particular power consumption and maintenance). Finally, MANTIS will present an opportunity to involve the South African science, engineering and industrial communities in this innovative technology in an early stage.

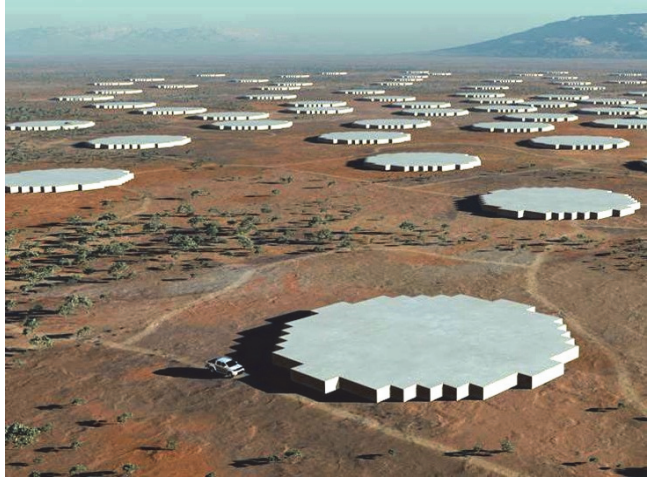
*Artist impression of the MFAA based SKA2 telescope. Credit: SKAO.*

# Introduction

The AAMID consortium has strong aspirations towards realizing the construction of a Mid-Frequency Aperture Array (MFAA) science demonstrator at the SKA South African site, co-located with MeerKAT and SKA1-Mid. It will be a first MFAA station on the SKA site and the effort can be gradually expanded as a major component of SKA Phase 2. At the same time, MANTIS will provide a reference point in terms of MFAA science, costing and technology performance towards an MFAA based SKA telescope.

Aperture arrays operating around 1 GHz have been successfully applied in technology demonstrators up to about 100 $m^2$ (the "EMBRACE" systems). Given the excellent results obtained with these technology demonstrators (Torchinsky et al. 2016), an MFAA based science instrument is a logical and essential next step. The aim is to realize this system on the South African SKA site. Consultations with the science community at the MFAA workshops in Stellenbosch (2014) and Cape Town (2016) indicated a strong interest, in particular in the areas of fast radio bursts (FRBs), HI intensity mapping, pulsars and SETI. In addition, the wide field of view (FoV) MANTIS instrument complements extremely well the very sensitive but narrower FoV MeerKAT and SKA1-Mid telescopes. The combination of the two offers unique science opportunities. For example, using the wide FoV of MANTIS, large areas of the sky can be permanently monitored for transient events. On occurrence, a trigger can be provided to MeerKAT for more sensitive and higher resolution follow-up observations. In the future, also the AVN and EVN can potentially extend the resolution of these follow-up observations to the milliarcsecond scale.

The multi-beaming, wide FoV and fast pointing capabilities of aperture arrays provide a very flexible system with many new modes of operation compared to existing dish-based instruments. Preparing for SKA2, MANTIS will allow scientists to explore how to optimally use these capabilities. This will result in very valuable input for the definition and design of the future large-scale MFAA systems, such as SKA2.

In parallel to the science observations, MANTIS will provide the required technology demonstration and validation. Several key aspects of MFAA technology can and will be demonstrated on a scale much smaller than a full station, such as sensitivity and power consumption. However, other key technologies require a system roughly the size of a full station to be validated, in particular the MFAA calibration concept and the verification of the station beam properties.

MANTIS will also provide a new reference for the costing of an MFAA based SKA, both in terms of deployment costs (hardware and installation) and operational costs (in particular power consumption and maintenance).

MANTIS will present an opportunity for the South African science and engineering community to get involved in this innovative technology at an early stage. Such involvement could and should include South African Universities, industries, and of course the SKA-SA engineering team. A first step in this direction, in the form of "educational tiles" placed at South African Universities, is already being prepared.

In this paper, we introduce the main characteristics of a wide-field SKA precursor that we envisage to be built at the South African SKA site, co-located with the MeerKAT and SKA1-Mid arrays. We review the current performance of MFAA technologies, and how we expect these to develop over the next few years. The main system specifications are considered in the light of four possible key science cases. The involvement of South Africa, and in particular South African industry and other centres of R&D expertise are also presented.

## MFAA Technology Development

Many of the most important discoveries in astronomy are the result of new technical innovations and growth (Harwit, 1981). Solla Price (1963) showed that the normal mode of growth is exponential, and any field that cannot maintain exponential growth will eventually die out. To maintain exponential growth, the continual introduction of new technology is required, since just refining or scaling up existing technology soon plateaus out (Ekers, 2010). Consequently, transformative technology is essential to maintain exponential progress in radio telescopes.

To continue exponential growth beyond 2025, at frequencies below 1.5 GHz, aperture arrays are the technology of choice. For the SKA, this allows dishes to be used for frequencies where they work best, i.e. $> 2 - 20$ GHz.

MFAA technology development has been steadily progressing over the last decade (Faulkner et al., 2010). Key technology challenges for MFAA are the reduction of power consumption, the reduction of capital and operational costs and the calibration of an MFAA system down to the thermal noise. In all these areas, major progress has been recorded. For example, Figure 1 shows that the costs and power consumption of antenna tiles have been reduced $2 - 4$ times over the last 5 years, and realistic projections for 2025 have also been made. Calibration strategies have been defined and a telescope-level optimization has been performed to balance costs between the front-end, CSP and SDP. Non-functioning prototypes have been placed at the Karoo site in 2014 to gain experience with long-term environmental effects and to provide input to the MFAA structural requirements.

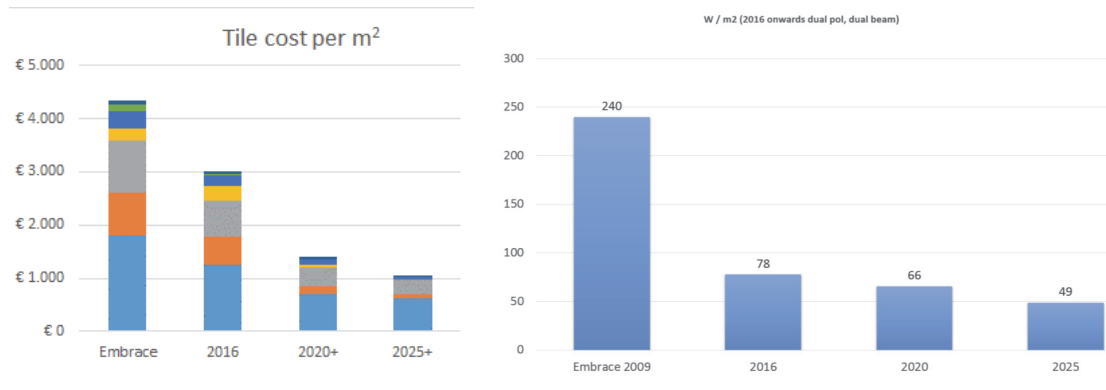

*Figure 1. The evolution and progress of Vivaldi tile costs per m² (top) and power consumption in W/m².*

Key aspects of MFAA technology (e.g. sensitivity and power consumption) can and will be demonstrated with engineering prototypes that are much smaller than MANTIS itself. However, several technologies require at least a full MFAA station to be validated. In particular, in order to demonstrate the MFAA calibration concept and verify the station beam pattern properties, a system of roughly the size of a single station is essential. The strategy to realize a sizable demonstrator has also been followed by LOFAR, MeerKAT and ASKAP (the "Initial Test Station", KAT-7 and BETA respectively) and has been extremely successful in mitigating risks and learning essential lessons before finalizing the design of the full system.

Obviously, there is a large synergy with LFAA, which is conceptually very similar to MFAA. We note that in the area of calibration and imaging, MANTIS will be able to take full advantage of the significant advances that have already been made with current instruments (e.g. Smirnov 2011). In particular, the analysis techniques that have enabled instruments such as LOFAR and the MWA to perform well in much less favorable ionospheric conditions, can also be adopted by MANTIS and SKA2.

## Preliminary Science Goals

MFAA technology will be essential to realize many of the SKA science goals (Braun, 2015). The MANTIS precursor seems particularly well suited for transient, pulsar, HI intensity mapping and SETI science.

### The Transient Radio Sky

A wide field instrument operating in the range 300—1400 MHz is ideally suited to the detection of highly impulsive transient phenomena whose radiation is generated by coherent emission mechanisms. Such emission is characterised by high ($T_b$ >> $10^{25}$K) brightness temperatures, short (t << 1s) durations and is often highly polarized. Fast Radio Bursts (FRBs) represent the most spectacular recent addition to this category. These objects are of extreme interest as cosmological probes of the baryonic content of the Inter-Galactic Medium, and they represent possible probes of the dark energy equation of state and the epoch of He reionization. Recent evidence appears to vindicate the hypothesis that at least some of these bursts do indeed emanate at cosmological distances (Lorimer et al. 2007; Keane et al. 2016). The ease with which the Parkes radio telescope detects these signals at (nominally) z ≤ 1.5 indicates that it may even be feasible to search for z > 6 bursts as probes of the epoch of H reionization.

The recent detection of gravitational wave events (Abbott et al. 2016) has lent impetus to attempts to detect their electromagnetic counterparts. The Advanced LIGO interferometers confine events to a region spanning > 600 deg² over an irregularly-shaped area. An aperture array possesses, in principle, the unique virtue of being able to span such a large and irregular FoV instantaneously. Events in the era of MANTIS should be confined to error regions $> \approx 3$ times smaller as further gravitational wave interferometers commence operation.

A search with a mid-frequency aperture array offers two advantages over existing and planned searches for transients at cosmological distances:

i) Frequencies > 1 GHz are necessary to detect high-redshift (z > 2) transients. Searches for FRBs at $\nu \sim 1$ GHz mitigate the deleterious effects of intergalactic dispersion ($\propto \nu^{-2}$) and scattering ($\propto \nu^{-4}$) which would otherwise hamper the detection of radiation from bursts from these distances.

ii) An aperture array possesses the flexibility to trade between sensitivity and instantaneous FoV. This attribute is extremely useful when one does not know how the event rate scales with burst flux density (i.e. the transients equivalent of source count statistics). For a population whose rate scales as $R \propto \Omega S_\nu^{-\gamma}$ (with γ=1.5 corresponding to homogeneously-distributed events in a Euclidean spacetime geometry) the optimal balance between sensitivity and FoV depends on the event rate index, γ. For instance, if we sub-divide the bandwidth by N in order to achieve a factor N increase in the number of beams, the event rate becomes $R \propto N^{(2-\gamma)/2} \Omega S_\nu^{-\gamma}$, so the greater number of beams is preferable when γ<2.

## Intensity Mapping

The earliest identified science driver for the SKA is the mapping of HI, including the measurement of the HI mass distribution on cosmological scales (see e.g. Rawlings et al. 2004). Indeed, an all-sky 'billion galaxy survey', using the redshifted HI 21-cm line, is at the heart of the SKA2 science case. By tracing the HI of individual galaxies over the entire sky and at different redshifts, SKA will be able to detect baryon acoustic oscillations (BAO), which constitute a preferred clustering scale in the distribution of matter on cosmological scales, and its evolution as a function of cosmic time (e.g. Abdalla et al. 2015; Bull et al. 2015). This will enable a direct measurement of the effects of dark energy (e.g. Rawlings et al. 2004; Abdalla & Rawlings 2005); in particular, below a redshift of 2, dark energy will start to dominate the Universe's expansion, and BAO detections will enable important distinctions between different cosmological models.

The method of intensity mapping (IM) has been proposed to make a statistical detection of the BAO in the HI mass distribution (e.g. Peterson et al. 2006; Ansari et al. 2008; Santos et al. 2015). It is likely to detect the first peak in the power spectrum, and a modest-sized precursor such as MANTIS can achieve this result. Moreover, HI emission observations are not affected by dust extinction, which may complicate competing optical/NIR projects such as the Euclid space mission (e.g. Laureijs 2012).

To significantly improve the precision of cosmological constants, a large volume needs to be sampled. For example, Euclid will survey 2 pi sr over a redshift range of 0.5 to 2.0, which corresponds to a co-moving volume of approximately 300 Gpc$^3$. MANTIS can also access (at least) this sky coverage, and it will also cover the same redshift range (corresponding to frequencies between about 500 to 1000 MHz). Thus, in principle it can access the same volume as Euclid.

IM observations with MANTIS requires a large FoV and a compact array configuration, in order to sample the typical scale of BAOs in one pointing and avoid complications from cosmic variance. The BAO scale peaks at ~ 100 Mpc/h which corresponds to about 1 deg (depending on the redshift). This means we need baselines between 10 m to 80 m in order to properly sample the BAO signal. The former will require flexibility at the sub-station level. Note that this also means we need a processed FoV of at least 5 x 5 deg$^2$ in order to view all the required scales in one pointing. As already pointed out, we will then need to survey a large area of the sky in order to have enough statistics (> 20,000 deg$^2$), so a larger FoV will be useful to increase the survey speed.

## Pulsars

Radio pulsars are rapidly-rotating neutron stars, which can be used as ultra-high-precision astronomical clocks for testing fundamental theories of gravity and dense matter physics. Timing measurements of a network of millisecond pulsars has the prospect of making the first detection of the gravitational wave background of the Universe in the nHz frequency range (e.g. Janssen et al. 2015). The discovery of 'holy grail' pulsar-black hole binary systems would allow completely novel tests of the basic characteristics of a black hole – mass, charge and spin – as well as the presence of an event horizon (e.g. Kramer et al. 2004). Moreover, neutron star mass measurements made via pulsar timing can also place unique constraints on the equation of state, for example in Demorest et al. (2010) where a measurement of a 2 solar mass neutron star was made via its Shapiro delay signature.

Increasing the number of known millisecond pulsars and relativistic pulsar binary systems is important, because only a small fraction of these systems provide competitive test-beds for studying fundamental physics (by virtue of their extreme orbital parameters, orbital geometry, masses, spin rates, etc.). Ultimately, a deep SKA all-sky survey is needed to find all the best 'laboratories' in our Galaxy. Pulsars have steep spectra, making them intrinsically brighter at lower frequencies, but scattering, dispersion, and sky temperature increase towards low frequency as well, mitigating the signal strength and effective time resolution. An optimum frequency range for pulsar surveys is therefore about 500-1400 MHz, with the higher frequencies being more appropriate for surveys close to and along the Galactic plane (e.g. Smits et al. 2009).

A primary goal of MANTIS is to demonstrate how a much larger mid-frequency aperture array system can perform a very deep pulsar survey, and also be a powerful instrument for pulsar timing. MANTIS will provide the equivalent collecting area of up to a 50-m dish (cf. the 64-m Parkes telescope), but with a FoV two orders of magnitude larger than facilities such as Parkes and MeerKAT (and up to a factor of a few larger than ASKAP). It will also not be subject to the detrimental effects of dish subarraying. Though the sensitivity of MANTIS is limited, known pulsars and new discoveries can be part of a continuous observing program as a proof of concept for the full SKA2-Mid timing observations. Given the wide FoV and multi-beam capability, this program can be run in parallel with other observations, or multiple pulsars can be timed simultaneously. For example, very-high-cadence monitoring could be used to search for starquake-induced glitches and single-pulse variations (e.g. Lyne et al. 2010), and there will also be novel applications for timing calibration.

## SETI

The Search for Extraterrestrial Intelligence (SETI) is a scientific research topic that has been prominently positioned as a key science goal of the SKA almost since the project's inception several decades ago. It goes without saying, that the detection of an artificial signal from an advanced civilization would have an enormous impact in many different areas, not least of all society at large. The question of whether or not we are alone in the Universe is one of the oldest questions that have engaged countless generations of mankind across the millennia.

Siemion et al. (2015) have recently detailed the capabilities of SKA1 and SKA2 with respect to detecting artificial radio signals from a variety of potential sources, including high-powered radar systems. For example, SKA2 should be capable of detecting planetary radar systems (such as those employed by Arecibo) for the nearest hundred thousand stars. MANTIS represents an important opportunity for SETI researchers – first of all it operates in the so-called "water hole" (the traditionally favored spectral window for SETI around 0.5-10 GHz), and secondly it opens up the possibility to survey large swathes of the sky simultaneously. The latter provides a new capability for SETI research at these frequencies, as current SETI instruments (e.g. Arecibo, GBT etc) provide only very restricted fields-of-view (albeit with high sensitivity). If SETI signals are very rare but also very strong, if they are also transitory in nature, MANTIS offers a new area of discovery space to explore.

For the most part, SETI observations performed by MANTIS could run commensally with other astronomical observations. Commensal SETI surveys would require a spigot to be in place that would provide a SETI backend with copies of the various beam formed data – a separate SETI backend is required in order to meet the extreme frequency (and time) resolution requirements of narrow-band signal searches (typically of order 1 Hz). Projects such as Breakthrough Listen might be interested in providing these back-end systems at limited cost to the MANTIS project. Nevertheless, we note that these systems also demand significant power, processing and archive capacity. In addition to this commensal approach, SETI researchers would also require modest amounts of targeted observing time for objects of special interest (e.g. the conjunction of extrasolar planetary systems along the Earth's line-of-sight).

Since the characteristics of SETI signals are currently unknown, it is difficult to be too prescriptive on the system requirements of this science case. From a sensitivity point of view, the preliminary goal of realizing a collecting area in excess of 1000 $m^2$ is probably a bare minimum in order to secure external investment. Since this is equivalent to a 25 metre telescope, MANTIS' niche is to offer a much larger FoV. This suggests a FoV requirement that is at least 10x better than the 25-m WSRT-APERTIF telescopes i.e. > 100 square degrees. Being able to simultaneously access the full front-end frequency band is also highly desirable.

# System Specifications
## MANTIS specifications

The specifications of MANTIS have not been finalized yet. However, the aim is to realize a system comparable to an SKA2 MFAA station, i.e. 1500 to 2500 $m^2$, equivalent to a 40 to 50 meter dish. Most of the collecting area is likely to be concentrated in one station, surrounded by a few smaller satellite stations. The latter will provide a test bed for interferometric observations. A few baselines beyond the core station are also required by some of the science cases.

Based on the prime science cases detailed in the next section, the preliminary specifications for MANTIS include:

- Frequency range: 450 - 1450 MHz
- Collecting area: 1500 – 2500 $m^2$
- A/T: 38 - 63 $m^2$/K
- SEFD: 74 - 44 Jy
- Field of view: 200 $deg^2$ at 1 GHz
- Bandwidth: >500 MHz
- Transient buffering

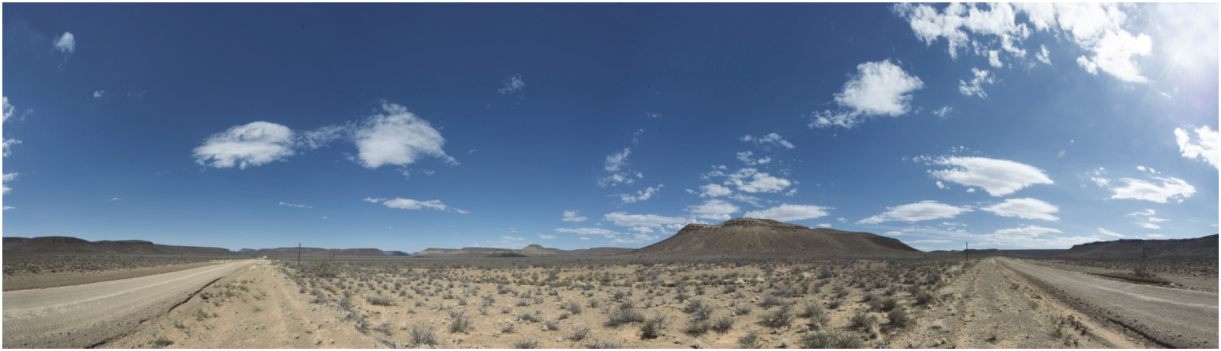

*Figure 2. Panoramic overview of the "K4" area at the Karoo site, a potential location of MANTIS (credit: Rob Millenaar).*

Based on the SKA site selection decision in 2012, the South African SKA site is the preferred site for the science demonstrator. SKA South Africa has identified a potential location, known as the "K4" area (see Figure 2). This site is surrounded by hills that offer excellent RFI shielding by terrestrial broadcasting and GSM, but also minimizes the risk of interfering with other activities at the site. As an SKA1-Mid dish is also planned in the "K4" area, there might be a possibility to use (or extend) some of the SKA1-Mid infrastructure.

## Comparison with other instruments

| | MANTIS | MeerKAT | JVLA | ASKAP | WSRT Apertif | UTMOST | CHIME | Tianlai (cylinder array) | Parkes Multibeam | Arecibo | FAST | SKA1-Mid |
|---|---|---|---|---|---|---|---|---|---|---|---|---|
| Operating frequencies (MHz) | 450-1450 | 700-10000 | 230-470, 1000-50000 | 700-1800 | 1110-1700 | 843 | 400-800 | 400-1420 | 440-24000 | 300-10000 | 100-3000 | 350-14000 |
| Bandwidth (MHz) | 500 | 1000 | 1000 | 300 | 300 | 31.25 | 400 | 400 | 400 | 1000 | 800 | 770 |
| Receptor size (m) | 40-50 | 13.5 | 25 | 12 | 25 | 778x11.6 | 100x20 | 120x15 | 64 | 225 | 300 | 15 |
| Fiducial frequency (MHz) | 1000 | 1400 | 1400 | 1400 | 1400 | 843 | 600 | 750 | 1400 | 1400 | 1400 | 1670 |
| FoV (deg$^2$) | 200 | 0.86 | 0.25 | 30 | 8 | 7.8 | 130 | 150 | 0.65 | 0.003 | 0.0017 | 0.49 |
| Maximum baseline (km) | 1 | 4 | 35 | 6 | 2.7 | 1.6 | 0.1 | 0.12 | 0.064 | 0.225 | 0.5 | 150 |
| Resolution (arcsec) | 60 | 11 | 1.3 | 7.4 | 16 | 45 | 1260 | 840 | 690 | 196 | 88 | 0.25 |
| A/T (m$^2$ K$^{-1}$) | 50 | 321 | 265 | 65 | 84 | 258 | 160 | 288 | 100 | 1150 | 1250 | 1560 |
| SEFD (Jy) | 60 | 8.6 | 10 | 42 | 33 | 11 | 17 | 9.6 | 28 | 2.4 | 2.2 | 1.8 |
| RMS (full instantaneous BW; µJy h$^{-1/2}$) | 32 | 3.2 | 3.9 | 29 | 22 | 32 | 10 | 5.6 | 16 | 0.89 | 0.92 | 0.75 |
| Example survey speed metric FoV x (A/T)$^2$ x BW (x 10$^8$ deg$^2$ m$^4$ K$^{-2}$ MHz) | 2.5 | 0.89 | 0.18 | 0.38 | 0.17 | 0.081 | 13 | 50 | 0.026 | 0.040 | 0.021 | 9.2 |
| Example normalized transients detection rate FoV x S$_0^{-3/2}$ (full instantaneous BW; flat spectral index assumed) | 1 | 0.13 | 0.029 | 0.17 | 0.067 | 0.038 | 3.6 | 9.9 | 0.0088 | 0.0032 | 0.0017 | 0.67 |

Notes: 100 per cent system efficiency assumed. UTMOST is a single-polarization instrument, and the survey speed value has therefore been reduced by a factor of 2.

References:
**MeerKAT, JVLA, ASKAP, Arecibo, Parkes Multibeam, FAST, SKA1-Mid**: SKA-TEL-SKO-0000002_03, SKA1 SYSTEM BASELINE DESIGN V2, Dewdney et al. 2016
**WSRT Apertif**: Oosterloo T. et al., 2010, ISKAF2010 Science Meeting, 43 (arXiv:1007.5141); http://www.ast.uct.ac.za/phiscc/sites/default/files/PHISCCtalks/Tom%20Oosterloo%20-%20Oosterloo.pdf
**UTMOST**: Caleb M. et al., 2016, MNRAS, 458, 718
**CHIME**: CHIME Collaboration 2012, CHIME Overview, http://chime.phas.ubc.ca/CHIME_overview.pdf; Bandura K. et al., 2014, Proc. SPIE, 9145, 914522; Bull P. et al., 2015, ApJ, 803, 21; Connor L.D., 2016, PhD thesis, U. Toronto
**Tianlai**: Xu Y. et al., 2015, ApJ, 798, 40; Bull P. et al., 2015, ApJ, 803, 21; Zhang J. et al., 2016, RAA, 16, 158

## South African Involvement

The development of the MANTIS precursor and an MFAA based SKA would also be very useful in employing and further exploiting South African industrial R&D expertise developed during the KAT-7 and MeerKAT program. Much of this resides in the SKA engineering office in Pinelands and in EMSS-Antennas (and also in Peralex, manufacturers of SKARAB) – there is particular expertise in DSP and RF & antenna engineering in these organizations. However, there is also expertise in radar engineering in Reutech Radar Systems and in high-tech satellite engineering and associated systems engineering (Space Advisory Company), all of who either participated or were represented at the recent MFAA workshop and also already involved in SKA-SA's Industry Participation program.

In support of research in support of the MFAA, a three-year EU FP-7 funded MIDPREP program, involving two European institutes (Chalmers, Sweden and ASTRON, the Netherlands) and three South African Partners (Stellenbosch University, Rhodes University, University of Cape Town) was put in place in mid-2013, and it has proven most valuable in developing South African human resources specifically for MFAA. South African students and researchers have become familiar faces at ASTRON and Chalmers in this period. To take this further, we are presently investigating the installation of small (one or two "tile") systems at SU, UCT or other possible sites, to further develop the skill set needed for both engineering design and scientific commissioning and exploitation of this MFAA technology.

This obviously connects very strongly with SKA-SA's highly successful Human Capital Development program, which over the past decade has supported several hundred undergraduate and postgraduate students, and the SKA-SA supported SARChI (SA Research Chair Initiative) chairs at SU, UCT, UWC, Wits, RU, and other chairs at in astronomy at UKZN and NWU.

## References

- Abdalla F.B., Rawlings S., 2005, MNRAS, 360, 27
- Abdalla F.B. et al., 2015, Proc. AASKA14, 017
- Ansari R. et al., 2008, arXiv:0807.3614
- Braun R. et al., 2015, Proc. AASKA14, 174
- Bull P. et al., 2015, Proc. AASKA14, 024
- Demorest P.B. et al., 2010, Nature, 467, 1081
- Ekers R.D., 2010, Proc. Of Science, arXiv:1004.4279
- Faulkner A., 2010, arXiv:1208.6180
- Harwit M., 1981, Cosmic Discovery, Basic Books Inc.
- Janssen G. et al., 2015, Proc. AASKA14, 037
- Kramer M. et al., 2004, New Astron. Rev., 48, 993
- Laureijs R. et al., 2012, Proc. SPIE, 8442, 84420T
- Lyne A. et al., 2010, Science, 329, 408
- Peterson J. et al., 2006, arXiv:astro-ph/0606104
- Rawlings S. et al., 2004, New Astron. Rev., 48, 1013
- Santos M. et al., 2015, Proc. AASKA14, 019
- Siemion, A. et al., 2015, AASKA14, 116 (arXiv:1412.4867).
- Smirnov O., 2011, A&A, 527, A110
- Smits R. et al., 2009, A&A, 493, 1161
- Solla Price D.J., 1963, Little Science, Big Science … and beyond, Columbia University Press
- Torchinsky S.A. et al, 2016, A&A, 589, A77, http://dx.doi.org/10.1051/0004-6361/201526706

## Affiliations

1. ASTRON, Netherlands Institute for Radio Astronomy, PO Box 2, 7990 AA, Dwingeloo, The Netherlands
2. Department of Physics and Astronomy, University of Western Cape, Cape Town 7535, South Africa
3. SKA SA, 3rd Floor, The Park, Park Road, Pinelands, 7405, South Africa
4. CENTRA, Instituto Superior Técnico, Universidade de Lisboa, Lisboa 1049-001, Portugal
5. ICRAR/Curtin University, Curtin Institute of Radio Astronomy, Perth, WA 6845, Australia
6. Department of Physics and Astronomy, University College London, Gower Place, London WC1E 6BT, U.K.

7. Department of Physics and Electronics, Rhodes University, PO Box 94, Grahamstown, 6140, South Africa
8. Dept. of Astronomy & Radio Astronomy Lab, Univ. of California, Berkeley, USA.
9. Dept. Of Astrophysics IMAPP, Radboud University, PO Box 9010, 6500 GL, Nijmegen, The Netherlands
10. Department of Astronomy, University of Cape Town, Private Bag X3, Rondebosch 7701, South Africa
11. Dept. Electric & Electronic Engineering, Univ. Stellenbosch, Stellenbosch, South Africa
12. Anton Pannekoek Institute for Astronomy, University of Amsterdam, PO Box 94249, 1090 GE Amsterdam, The Netherlands
13. Jodrell Bank Centre for Astrophysics, School of Physics & Astronomy, The University of Manchester, Alan Turing Building, Oxford Road, Manchester, M13 9PL, UK.
14. Leiden Observatory, Leiden University, PO Box 9513, 2300 RA Leiden, The Netherlands.
15. Cavendish Laboratory, Department of Physics, University of Cambridge, Cambridge CB3 0HE, UK
16. Station de radioastronomie de Nançay, Observatoire de Paris, CNRS, France
17. Joint Institute for VLBI ERIC, PO Box 2, 7990 AA, Dwingeloo, The Netherlands

# List of Abbreviations

| | |
|---|---|
| AAMID | Aperture Array MID frequency consortium |
| ASKAP | Australian SKA Pathfinder |
| AVN | African VLBI Network |
| BAO | Baryonic Acoustic Oscillations |
| BETA | Boolardy Engineering Test Array |
| CDR | Critical Design Review |
| CSP | Central Signal Processor |
| EU | European Union |
| EVN | European VLBI Network |
| FoV | Field of View |
| FP7 | Seventh Framework Programme |
| FRB | Fast Radio Burst |
| GSM | Global System for Mobile Communications |
| INFRA-SA | Infrastructure Africa |
| IM | Intensity Mapping |
| KAT | Karoo Array Telescope |
| LFAA | Low Frequency Aperture Array |
| LNA | Low Noise Amplifier |
| LOFAR | Low Frequency Aperture Array |
| MANTIS | Mid-Frequency Aperture Array Transient and Intensity-Mapping System |
| MIDPREP | Preparing for MID-SKA receiving concepts in South Africa |
| MFAA | Mid Frequency Aperture Array |
| MWA | Murchison Widefield array |
| NWU | North-West University |
| PDR | Preliminary Design Review |
| RF | Radio Frequency |
| RFI | Radio Frequency Interference |
| RU | Rhodes University |
| SA | South Africa |
| SDP | Signal Data Processing |
| SEFD | Source Equivalent Flux Density |
| SETI | Search for Extraterrestrial Intelligence |
| SDP | Science Data Processor |
| SKA | Square Kilometre Array |
| SKAO | SKA Office |
| SKA-SA | SKA South Africa |
| SU | Stellenbosch University |
| TBD | To Be Determined |
| UCT | University of Cape Town |
| UKZN | University of KwaZulu-Natal |
| UWC | University of Western Cape |
| Wits | University of the Witwatersrand |